\documentclass[twocolumn,showpacs,preprintnumbers,amsmath,amssymb,prl,superscriptaddress]{revtex4}

\usepackage{amsfonts}
\usepackage{graphicx}
\usepackage{dcolumn}
\usepackage{bm}

\begin{document}

\preprint{APS/123-QED}

\title{Enhanced quasiparticle heat conduction of the multigap superconductor Lu$_{2}$Fe$_{3}$Si$_{5}$}
% \title{Low energy quasiparticle excitations in multigap superconductor Lu$_{2}$Fe$_{3}$Si$_{5}$\\: Thermal conductivity study}

\author{Y. Machida}
\affiliation{Department of Physics, Tokyo Institute of Technology, Meguro, Tokyo 152-8551, Japan}
\author{S. Sakai}
\affiliation{Department of Physics, Tokyo Institute of Technology, Meguro, Tokyo 152-8551, Japan}
\author{K. Izawa}
\affiliation{Department of Physics, Tokyo Institute of Technology, Meguro, Tokyo 152-8551, Japan}
\author{H. Okuyama}
\affiliation{Department of Physics, College of Science and Technology, Nihon University, Chiyoda, Tokyo 101-8308, Japan}
\author{T. Watanabe}
\affiliation{Department of Physics, College of Science and Technology, Nihon University, Chiyoda, Tokyo 101-8308, Japan}

\date{\today}

\begin{abstract}
The thermal transport measurements have been made on the Fe-based superconductor Lu$_{2}$Fe$_{3}$Si$_{5}$ ($T_c \sim 6$ K) down to a
very low temperature $T_c$/120.
The field and temperature dependences of the thermal conductivity confirm the multigap superconductivity with
fully opened gaps on the whole Fermi surfaces.
In comparison to MgB$_2$ as a typical example of the multigap superconductor in a $p$-electron system, 
Lu$_{2}$Fe$_{3}$Si$_{5}$ reveals a remarkably enhanced quasiparticle heat conduction in the mixed state.
The results can be interpreted as a consequence of the electronic correlations derived from Fe 3$d$-electrons.
% Thermal conductivity $\kappa$ of single crystalline Lu$_{2}$Fe$_{3}$Si$_{5}$ has been measured
% down to 0.05 K and under magnetic field up to 7 T. 
% A residual $T$-linear term of $\kappa(T)$ in the zero temperature
% limit is found to be negligibly small pointing to a fully gapped $s$-wave superconductivity.
% In the mixed state, $\kappa(H)$ reveals a rapid increase and reaches a half of the normal state value under low field $\sim 0.05 H_{\rm c2}$.
% The observed behavior is interpreted in terms of multigap effects.
% The detail analysis of $\kappa(T)$ under fields based on the two-gap model provides field variations of the superconducting gaps.
\end{abstract}

% \pacs{75.47.-m, 72.15.-v, 75.30.Mb}

\maketitle
Multigap superconductivity (MGSC) is the existence of superconducting gaps with significantly different magnitude
on distinct Fermi surfaces.
This phenomena has been realized in wide variety of materials
including MgB$_2$~\cite{MgB2specific}, NbSe$_2$~\cite{NbSe2} and the heavy fermions~\cite{Pr,URu2Si2,Ce},
pointing to its universality underlying the superconductivity.
% This unusual phenomenon has recently emerged as a possible explanation for the anomalous properties
% of some superconductors.
One consequence of the MGSC is the ability to
excite low-energy quasiparticles (QPs) due to the presence of the small gap, 
providing unusual features in the mixed state.
Even though the MGSC has been extensively studied so far,
the purpose of the study has been mainly focused on the explanation 
for the anomalous properties observed in the superconducting state.
Recent discovery of the iron-pnictide superconductors~\cite{kamihara}, however,
have offered further insight into the MGSC especially in the
correlated electron system
because the electronic interactions and the multiband structure are essential for the pairing mechanism in this system~\cite{ishida}.
In that sense, a question of how the electronic correlations
affect on the MGSC is a fascinating issue to be addressed,
which could not be examined on MgB$_2$.
Unfortunately, lack of high-quality stoichiometric samples and/or the high upper critical field
prevent detailed studies of the MGSC in the iron-pnictides~\cite{ishida}.
% Since the iron-pnictide superconductors were discovered, however,
% the detail clarification of the MGSC
% has been strongly required,
%  for understanding the superconducting properties, 
% Moreover, 
% a question of how the electronic correlations derived from the Fe 3$d$-electrons
% affect on the MGSC is a fascinating issue to address, 
% a relation between the electronic correlations derived from the Fe 3$d$-electrons and the MGSC
% has remained a fascinating problem to be solved.
% which can make the properties of this $s$-wave superconductor similar to those of $d$-wave superconductors. 
% Among the multigap superconductors, MgB$_2$ is a well established case
% by means of the specific heat measurements, photoemission spectroscopy and so on~\cite{MgB2, MgB2specific,MgB2ARPES}.
% Recently, the multigap superconductivity is found not only in conventional superconductors but also in 
% unconventional one including the heavy fermion SC~\cite{Pr,Ce}.

% Multigap superconductivity in Lu$_{2}$Fe$_{3}$Si$_{5}$~\cite{nakajima,gordon,josephson}.
Lu$_{2}$Fe$_{3}$Si$_{5}$, a ternary-iron-silicide,
is an another example of the Fe-based multigap superconductor with $T_c \sim 6$ K~\cite{nakajima}, 
which crystallizes in the tetragonal structure consisting
of a quasi-one-dimensional iron chain along the $c$
axis and quasi-two-dimensional iron squares parallel to the basal plane~\cite{Braun}.
Band structure calculations predict that 
Fermi surfaces consist of
two holelike bands and one electronlike band, and
each band has a contribution from the Fe 3$d$-electrons~\cite{nakajima}.
In the two holelike bands, some parts of the Fermi surfaces have the different dimensionality reflecting the crystal structure~\cite{nakajima}.
Importantly, the Fe 3$d$-electrons are responsible for the superconductivity of Lu$_{2}$Fe$_{3}$Si$_{5}$,
as suggested by the absence of superconductivity in the isoelectronic Lu$_{2}$Ru$_{3}$Si$_{5}$ and
Lu$_{2}$Os$_{3}$Si$_{5}$~\cite{Braun2}.
Therefore, Lu$_{2}$Fe$_{3}$Si$_{5}$ stands as the rare stoichiometric multigap superconductor with the $d$-electrons in between 
the $p$-electron system (e.g., MgB$_2$~\cite{MgB2specific}) 
and the $f$-electron system such as PrOs$_4$Sb$_{12}$~\cite{Pr},
providing an unique opportunity to study the MGSC in the moderately correlated electron system.

The multigap superconductivity of Lu$_{2}$Fe$_{3}$Si$_{5}$ is first observed by the 
specific heat measurement down to 0.3 K under zero field~\cite{nakajima}.
It is of interest to further elucidate
the MGSC by performing
the thermal conductivity measurements down to lower temperature in the vortex state,
because the thermal conductivity is sensitive to the light carrier band 
which is expected to strongly affect on the low-energy QP excitations.
Absence of the nuclear schottky contribution is another advantage of this technique
in exploring the MGSC.
% The advantage of Lu$_{2}$Fe$_{3}$Si$_{5}$ for studying the multigap superconductivity is 
% (1) the ability of the superconductivity without the impurity doping, 
% (2) the low accessible upper critical field, and
% (3) the simple band structure.
% The Josephson effect suggested
% the spin-singlet superconductivity in Lu$_{2}$Fe$_{3}$Si$_{5}$~\cite{josephson}.

In this paper, we report on 
the thermal transport measurements of single crystalline Lu$_{2}$Fe$_{3}$Si$_{5}$ 
down to $T_c$/120.
Our detailed results of the thermal conductivity in the mixed state confirm the MGSC in Lu$_{2}$Fe$_{3}$Si$_{5}$.
% A residual $T$-linear term of the thermal conductivity in the zero temperature
% limit is found to be negligibly small pointing to a fully gapped superconductivity.
% In the mixed state, the heat transport probes the delocalized quasiparticles even under low fields $H < H_{c2}/25$
% in striking contrast with a fully gapped $s$-wave superconductor.
% $\kappa(H)$ reveals a rapid increase and reaches a half of the normal state value under low field $\sim 0.05 H_{\rm c2}$.
% In addition, the detail analysis of the thermal conductivity in the vortex state based on the two-gap model
% provides the field variations of the superconducting gaps,
% reminiscence of the two-gap superconductor with a weak interband interaction.
% These results point to the multigap superconductivity in Lu$_{2}$Fe$_{3}$Si$_{5}$.
Moreover, from a comparative study with MgB$_2$,
Lu$_{2}$Fe$_{3}$Si$_{5}$ reveals the significantly enhanced heat conduction
as an indication of the electronic interactions derived from the Fe 3$d$-electrons.

Single crystals
were grown by the floating-zone method~\cite{watanabe}.
One-heater-two-thermometer steady-state method was used to measure the thermal conductivity.
The heat current $q$ was aligned along the [001] direction, 
and the magnetic field is applied parallel to the $ab$ plane. 
% We used Cernox and RuO$_2$ thermometers above and below 1 K, respectively.
The thermal contacts with resistance of $\sim$10 m$\Omega$
were made to the sample by using a spot welding technique.
% The thermometers were thermalized on the sample by gold wires held by spot welding,
% providing a good thermal contact with low contact resistance $R_c \leq$ 10 m$\Omega$ at 300 K.
% The same contacts and gold wires were used
% to measure the resistivity by a
% standard four-contact method.

\begin{figure}[t]
\begin{center}
\includegraphics[scale =0.45]{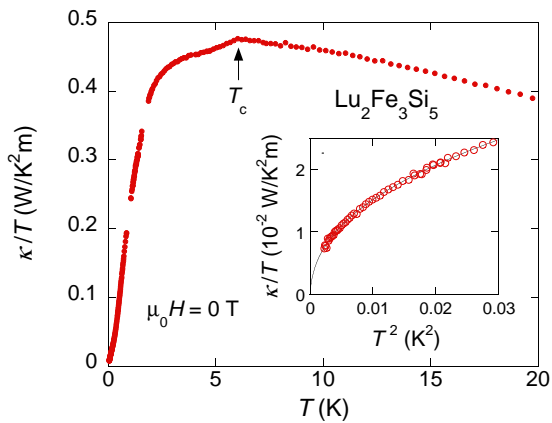}
\end{center}
\vspace{-0.5cm}
\caption{\label{fig.1} (color online). Temperature dependence of the thermal conductivity divided by temperature $\kappa/T$ 
under zero field.
The arrow denotes the superconducting temperature $T_c$. 
Inset: $\kappa/T$ vs. $T^2$ plot under zero field. 
The solid line represents a fit to the data by $\kappa/T=\kappa_{\rm 0}/T + b T^{\alpha-1}$.}
\end{figure}

Figure~\ref{fig.1} shows the temperature dependence of the thermal conductivity divided by temperature $\kappa(T)/T$
under zero field.
An arrow denotes the superconducting transition temperature $T_{\rm c} \sim$ 6 K.
% which corresponds to the temperature where the resistivity becomes zero as shown in the inset of Fig.~\ref{fig.1}.
% The Lorentz number $L=\kappa\rho/T=1.4L_0$ at $T_{\rm c}$ is larger than the Sommerfeld value 
% $L_0=2.44\times10^{-8}$ $\Omega$W/K$^2$, indicating a phonon contribution in the heat conduction near $T_{\rm c}$.
As clearly seen, $\kappa(T)/T$ shows a kink at $T_{\rm c}$ followed by steep decrease with decreasing temperature.
On further cooling, a hump structure appears around 3 K.
%The inset of Fig.~\ref{fig.1} shows the temperature dependence of the resistivity under zero field.
%A narrow superconducting transition width $\Delta T_{\rm c} \leq 0.2$ K indicates the high quality of our single crystals.
The decrease of $\kappa(T)/T$ in the superconducting state
attributes to the reduction of the QP density by opening the superconducting gaps.
% as has been observed in several superconductors.
On the other hand, the hump structure originates from an enhanced
phonon mean free path due to the condensation of electronic
scattering centers in the superconducting state.
In fact, similar enhancement of the phononic conductance (called phonon peak) in the superconducting state
is also observed in various materials such as Nb~\cite{Nb1} and Pb~\cite{Pb}.
The inset of Fig.~\ref{fig.1} shows $\kappa/T$ vs. $T^2$ plot.
Since the measured thermal conductivity contains both electron and phonon contributions,
% Since we measure the total thermal conductivity, 
it is necessary to separate to each part as $\kappa=\kappa_{\rm e}+\kappa_{\rm ph}$.
Below 0.15 K, $\kappa/T$ is well described by a relation of $\kappa/T = \kappa_0/T + b T^{\alpha-1}$ 
with $\kappa_0/T=0.000\pm0.005$W/K$^2$m, $b$ = 0.11 W/K$^{2.8}$m, and $\alpha \sim$ 1.8.
Here, we should note that the conventional $T^3$ dependence of the phonon thermal conductivity $\kappa_{\rm ph}$
dominated by the ballistic phonon scattering fails to reproduce our result.
% This enables a separation of the electronic and the phononic thermal conductivities.
The lower power of $\kappa_{\rm ph}$ with $\alpha \sim 1.8$ 
can be attributed to either the specular reflection of the phonon or the phonon scattering off the electrons~\cite{phonon1,phonon2}. 
For the electronic contribution, $\kappa_0/T$ provides the electronic thermal conductivity in the zero temperature limit.
The negligibly small $\kappa_0/T$ clearly indicates that Lu$_{2}$Fe$_{3}$Si$_{5}$ is a fully gapped superconductor
consistent with the specific heat measurement~\cite{nakajima}.

\begin{figure}
\begin{center}
\includegraphics[scale =0.48]{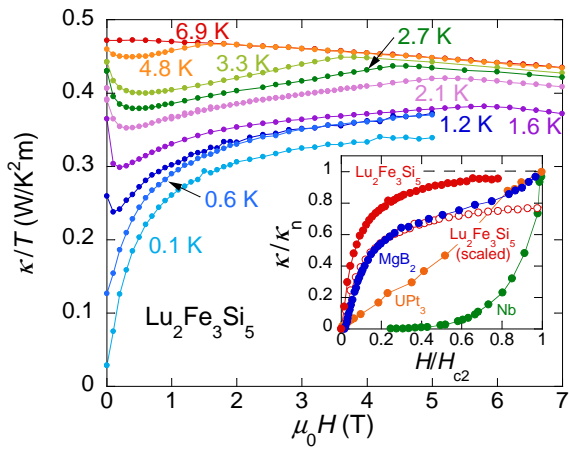}
\end{center}
\vspace{-0.5cm}
\caption{\label{fig.2} (color online). Field dependence of the thermal conductivity divided by temperature $\kappa/T$ 
at several temperatures.
Inset: The thermal conductivity normalized to its normal state value $\kappa/\kappa_{\rm n}$ vs. $H/H_{\rm c2}$ plot at 0.1 K. 
For comparison, the data for Nb~\cite{Nb2}, MgB$_2$~\cite{mgb2thermal}, and UPt$_3$~\cite{UPt3} are also shown.
The open circles show a result of scaling described in the text.}
\end{figure}

Figure~\ref{fig.2} shows the field dependence of the thermal conductivity divided by temperature $\kappa(H)/T$ 
at several temperatures.
For $T \geq$ 1.0 K, $\kappa(H)/T$ takes a minimum at low fields and then increases with field.
% The value of $H_{\rm c2}$ is determined as a maximum in $\kappa(H)/T$.
% The depth of a minimum as well as the sharpness of the drop become
% more pronounced as temperature is lowered.
% Similar behavior in the mixed state has also been observed on Nb~\cite{Nb1}.
In general, the existence of the minimum of $\kappa(H)/T$ is explained as the result of
the decrease of the phonon conductivity due to the vortex scattering concomitant with the increase of the 
electronic conductivity resulting from the increase of the
delocalized QP density~\cite{Nb2,mgb2thermal,NbSe2}. 
%The drop of $\kappa/T$ is originated from decrease in the phonon thermal conductivity.
Remarkably, for $T <$ 1.0 K, $\kappa(H)/T$ shows rapid increase and reaches almost half of the normal state value $\kappa_{\rm n}/T$ 
already at low fields ($\mu_0H <$ 1 T).
Here, $\kappa_{\rm n}/T$ is estimated from the Wiedemann-Franz law via
$\kappa_{\rm n}/T=L_0/\rho_0$ = 0.325 W/K$^2$m,
where $L_0=2.44 \times 10^{-8} {\rm W}\Omega/{\rm K}^2$ is the Lorenz number,
and $\rho_0$ = 7.5 $\mu\Omega$cm is the normal-state residual resistivity.
% obtained
% by the extrapolation of normal state resistivity by assuming $\rho=\rho_0+AT^2$.
% $\mu_0H$ = 0.4 T ($\sim$ 0.05 $H_{\rm c2}$).
% One immediately notices the large amount of delocalized quasiparticles
% throughout the vortex state of Lu$_{2}$Fe$_{3}$Si$_{5}$.
It is of interest to compare our results with several superconductors.
The inset of Fig.~\ref{fig.2} depicts
the normalized thermal conductivity $\kappa/\kappa_{\rm n}$
at 0.1 K plotted against $H/H_{\rm c2}$
together with the data for Nb~\cite{Nb2}, MgB$_2$~\cite{mgb2thermal}, and UPt$_3$~\cite{UPt3}.
Here, the phonon thermal conductivity $\kappa(H= 0)$
is subtracted from $\kappa/\kappa_{\rm n}$, and the upper critical field of $\mu_0H_{\rm c2}$ = 6.4 T is obtained from Ref.~\cite{nakajima2}.
The field variation of $\kappa/\kappa_{\rm n}$ for Lu$_{2}$Fe$_{3}$Si$_{5}$ is in dramatic contrast with the behavior of the
fully gapped $s$-wave superconductor Nb~\cite{Nb2}, 
in which small magnetic fields hardly affect the low temperature
thermal conductivity. 
By contrast, in nodal
superconductor UPt$_3$,
the delocalized QPs induced by the Doppler shift produces 
the remarkable field dependence of $\kappa/\kappa_{\rm n}$.
% Indeed, half of the normal
% state thermal conductivity is restored already at H $\approx$ 0.05$H_{\rm c2}$ for MgB$_2$, 
% and about 40$\%$ of $\kappa$ 
% at H $\approx$ 0.05$H_{\rm c2}$ in the case of PrOs$_4$Sb$_{12}$.
The strongly enhanced thermal conductivity of Lu$_{2}$Fe$_{3}$Si$_{5}$ is a clear indication of either a gap with nodes
or nodeless multiple gaps with significantly different magnitude on distinct Fermi surfaces.
However, the nodal gap is ruled out by the absence of the $T$-linear term of $\kappa(T)$.
% Strong evidence for multigap superconductivity is provided by the field dependence of the thermal conductivity.
% For comparison, $\kappa/\kappa_{\rm n}$ for ,
% are also plotted.
% MgB$_2$ is a well established case of multigap superconductivity with a small gap on one Fermi surface and
% a larger gap on the other~\cite{MgB2thermal}.

On the other hand, one immediately notices that Lu$_{2}$Fe$_{3}$Si$_{5}$
shares striking resemblance with MgB$_2$, namely,
a rapid increase of $\kappa/\kappa_{n}$ at low fields and a saturation behavior at high fields,
although $\kappa/\kappa_{n}$ for Lu$_{2}$Fe$_{3}$Si$_{5}$ shows even more pronounced increase and takes higher values.
% Here, we define $H_{c2}^{s}\sim 0.03H_{c2}$ for Lu$_{2}$Fe$_{3}$Si$_{5}$ and
% $\sim 0.1H_{c2}$ for MgB$_2$, respectively,  as a field above which $\kappa/\kappa_{n}$ starts to level off
% as denoted by the arrows in the inset of Fig.~\ref{fig.2}.
% i.e. distinct shoulders manifest around $H_{\rm c2}/10$ in both curves.
The field evolution of the thermal conductivity of MgB$_2$
is well understood in terms of the multigap superconductivity with a small gap $\Delta_{s}$ on one Fermi surface 
and a large gap $\Delta_{l}$ on the other (index $l$ and $s$ represent large and small gaps, respectively)~\cite{mgb2thermal}.
A consequence of the small gap is an existence of a ``virtual upper critical field" $H_{c2}^{s}$,
above which the overlap of huge vortex core provides a dramatic increase of the delocalized QPs contributed to the heat conduction.

% The fast growth of $\kappa/\kappa_{n}$ originates from the field-induced suppression of the smaller gap $\Delta_{s}$
% and resultant increase of delocalized QPs excited above $\Delta_{s}$.
% With this regards, the characteristic field scale of $H_{c2}^{s}$ is interpreted as a ``virtual upper critical field".
% The shoulder-like structure indicates the saturation of the contribution of the small gap band carriers.
% because the quasiparticles are easily excited over the field-suppressed $\Delta_{s}$ above $H_{c2}^{s}$.
The characteristic field scale of $H_{c2}^{s}$ can be estimated from 
$H_{c2}^{s}/H_{c2} \sim (\Delta_{s}/\Delta_{l})^2(v_{F,l}/v_{F,s})^2$.
On the other hand, the ``normal state" contribution of the small gap band $\kappa_{s}/\kappa_{n}$ is
obtained from $\kappa_{s}/\kappa_{n}=\kappa_s/(\kappa_{l}+\kappa_{s})=N_sv_{F,s}^2\tau_s/(N_lv_{F,l}^2\tau_l+N_sv_{F,s}^2\tau_s)$
via a relation of $\kappa_i \propto N_iv_{F,i}^2\tau_i$, where $N_i$, $v_{F,i}$, and $\tau_i$ represent the normal state electronic densities of state, 
the Fermi velocity, and the scattering rate of each gap band ($i=l,s$).
Here, suppose $\tau_{s}/\tau_{l}$ is close to unity, $\kappa_{s}/\kappa_{n}$ is simplified to $\sim 1/\{1+(N_l/N_s)(v_{F,l}/v_{F,s})^2\}$.
Consequently, the two gap superconductivity is characterized by 
the three ratios $\Delta_{l}/\Delta_{s}$, $N_l/N_s$, and $v_{F,l}/v_{F,s}$.
With the knowledge of $\Delta_{l}/\Delta_{s} \sim 4$~\cite{MgB2specific} and  
$N_l/N_s$, $v_{F,l}/v_{F,s} \sim 1$~\cite{mgb2thermal}, 
we find $H_{c2}^{s}/H_{c2}\sim$ 1 and $\kappa_{s}/\kappa_{n}\sim$ 0.5 for MgB$_2$
in agreement with various estimations~\cite{mgb2thermal,bouquet3}.
Now, let us discuss the case of Lu$_{2}$Fe$_{3}$Si$_{5}$.
Since $\Delta_{l}/\Delta_{s} \sim 4$ and $N_l/N_s \sim 1$~\cite{nakajima} as same as MgB$_2$,
the $\kappa/\kappa_{n}$ curve of Lu$_{2}$Fe$_{3}$Si$_{5}$ is expected to be scaled to that of MgB$_2$ by tuning the ratio of $v_{F,l}/v_{F,s}$.
We achieve excellent scaling with $v_{F,l}/v_{F,s}\sim$ 0.8 as shown by th open circles in the inset of Fig.~\ref{fig.2},
yielding $H_{c2}^{s}/H_{c2}\sim$ 0.04 and $\kappa_{s}/\kappa_{n}\sim$ 0.6.
The deviation at high fields is due to the difference of $H_{c2}$ in these two systems. 
% For Lu$_{2}$Fe$_{3}$Si$_{5}$,
% we reproduce $H_{c2}^{s}/H_{c2} \sim 0.03$ with $v_{F,l}/v_{F,s} \sim 0.7$
% by assuming that the small gap opens on the light carrier band.
% Here, we use the gap ratio of  $\Delta_{l}/\Delta_{s} = 4$, which is determined by the specific heat measurement~\cite{nakajima}.
% To verify the assumption of $v_{F,l}/v_{F,s} < 1$,
% we alternatively estimate the ratio of $v_{F,l}/v_{F,s}$ from 
% the value of $\kappa/\kappa_{n}(=\kappa_{s}/(\kappa_{l}+\kappa_{s}) \sim 0.8)$ at the shoulder
% Suppose $\tau_{s}/\tau_{l}$ is close to unity,
% $\kappa/\kappa_{n}$ is reduced to 
% $\sim v_{F,s}^2/(v_{F,s}^2 + v_{F,l}^2)$,
% using $N_{s}/N_{l} \sim 1$ as indicated by the specific heat measurement~\cite{nakajima}.
% We then get $v_{F,l}/v_{F,s} \sim 0.5$, which is close to the estimated value from $H_{c2}^{s}$.
The inequality of $v_{F,l}/v_{F,s}$ indicates that the carrier mass $m$, which is inversely proportional to $v_{\rm F}$ ($m \propto1/v_{\rm F}$), is different in each band
for Lu$_{2}$Fe$_{3}$Si$_{5}$ in contrast to MgB$_2$~\cite{mgb2thermal}.
This contradiction might originate 
from the character of the dominant electrons contribute to the density of state at the Fermi surface;
Fe 3$d$-electrons in Lu$_{2}$Fe$_{3}$Si$_{5}$ and B 2$p$-electrons in MgB$_2$, respectively.
In addition, unequal weight of the $d$ character among the distinct Fermi surfaces yields the heavy and light carrier bands in Lu$_{2}$Fe$_{3}$Si$_{5}$.
Notably, a signature of the electronic correlations possibly derived from the $d$-electrons
can be found in
%  may result in the heavy carrier band with large gap in Lu$_{2}$Fe$_{3}$Si$_{5}$.
% from the Fe 3$d$-electron correlations in Lu$_{2}$Fe$_{3}$Si$_{5}$.
the specific heat coefficient $\gamma_n=23.7$ mJ/molK$^2$~\cite{nakajima}, which is about 10 times
larger than that of MgB$_2$ ($\gamma_n=2.62$ mJ/molK$^2$)~\cite{Choi}.
The rather small $\gamma_{\rm band}=8.69$ mJ/molK$^2$ obtained from the band calculations~\cite{nakajima} also indicates
the presence of the electronic interactions.
On the other hand, the small gap on the light carrier band provides
dramatic enhancement of $\kappa(H)/T$  in the mixed state than that of MgB$_2$
because $\kappa$ is sensitive to the light carrier mass.

\begin{figure}
\begin{center}
\includegraphics[scale =0.45]{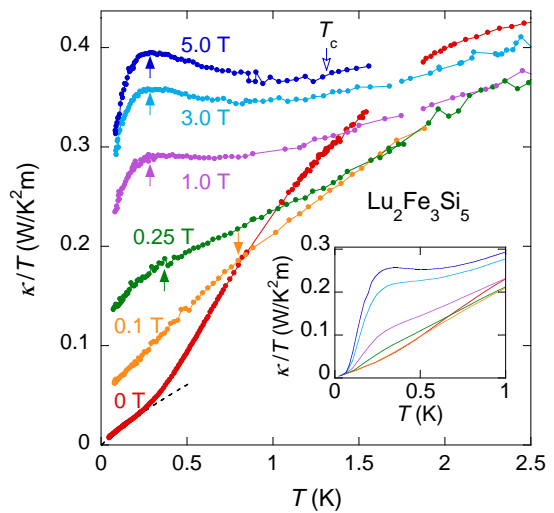}
\end{center}
\vspace{-0.5cm}
\caption{\label{fig.3} (color online). Temperature dependence of the thermal conductivity divided by temperature $\kappa/T$ 
under several magnetic fields.
The solid and open arrows denote the anomalies associated with the small gap and $T_c$ at 5 T, respectively.
The dashed line represents a fit to the data by $\kappa/T=\kappa_{\rm 0}/T + b T^{\alpha-1}$.
Inset: The calculated $\kappa/T$ curves for the corresponding fields within the two-gap model
described in the text.}
\end{figure}

In order to further clarify the multigap superconductivity of Lu$_{2}$Fe$_{3}$Si$_{5}$,
we present the temperature dependence of $\kappa(T)/T$ in the vortex state (Fig.~\ref{fig.3}).
% Figure~\ref{fig.3} shows the temperature dependence of the thermal conductivity $\kappa/T$
% under several fields.
By applying magnetic fields, we
observe a pronounced increase of $\kappa(T)/T$ at low temperatures 
corresponding to the fast growth of $\kappa(H)/T$ at low fields.
Furthermore, an anomaly associated with a slight change of slope is found around 0.8 K at 0.1 T,
and it shifts to $\sim 0.3$ K at 0.25 T as denoted by the arrows in Fig.~\ref{fig.3}. 
% $\kappa/T$ rises to half of the normal state $\kappa/T$ at 0.1 K under small field $\mu_0H=0.25$ T.
With further increasing fields, the anomaly becomes more pronounced, being a maximum
around 0.2 K above 1.0 T.
The emergence of the maximum can be attributed to either the phononic or
the electronic contribution.
% We attribute the origin of the maximum to
% an enhancement of the quasiparticle mean free path $l_e$, (3) a magnetic anomaly associated with a magnetic order, and
% (4) . 
However, we immediately exclude the phononic origin because the appearance of phonon peaks twice below $T_c$ is
highly unlikely.
Moreover, 
an increase of the phonon mean-free path up to 10 folds of the zero-field value, for example at 5 T,
% an increase of the phonon mean-free path up to 10 times lager than the zero-field value compared with the one at 5 T
is improbable even if the low-temperature thermal conductivity is dominated by the specular reflection.
Thus, we turn to the electronic origin raising the following possibilities;
(1) an enhancement of the QP mean free path $l_e$, (2) a magnetic anomaly associated with a magnetic order, and
(3) an increase of delocalized QPs excited above the field-suppressed small gap $\Delta_{s}$.
For the possibility of (1),
the enhancement of $l_e$ upon entering the superconducting state
usually occurs as a result of the strong inelastic scattering.
% The quasiparticle mean-free path in the normal state is strongly suppressed
% by the inelastic scattering due to antiferromagnetic fluctuations.
% On the other hand, in the superconducting state the electrons are condensed into Cooper pairs 
% and their numbers decrease rapidly below $T_{\rm c}$. 
% This gives rise to a reduction of the scattering cross section of QPs, 
% and hence $l$ increases below $T_{\rm c}$.
This is because
normal state $l_e$ suppressed by the
inelastic scattering starts to increase in the superconducting state
due to the condensation of electronic scattering centers.
In practice, this behavior has been observed in CeCoIn$_5$~\cite{kasahara} and YBCO~\cite{YBCO},
in which a source of the inelastic scattering is attributed to the magnetic fluctuations.
However, no signature of such magnetic fluctuation has been indicated in Lu$_{2}$Fe$_{3}$Si$_{5}$~\cite{mag}.
The magnetic origin of the maximum is also ruled out because Fe atoms carry no magnetic moment~\cite{mag}.
Therefore, in the following we discuss the remaining possibility (3) by analyzing $\kappa(T)/T$ within the framework of the two-gap model.

In the superconducting state, the electronic thermal conductivity $\kappa_e$ is expressed as, 
\begin{equation}
\kappa_e/\kappa_n=\frac{2F_1(-y)+2y{\rm ln}(1+e^{-y})+y^2(1+e^y)^{-1}}{2F_1(0)},
\label{eq.1}
\end{equation}
where $y=\Delta(t)/k_{\rm B}T$, $\Delta(t)$ being the half of the energy gap
and $t=T/T_c$~\cite{BRT}.
The temperature dependence of $\Delta(t)$ is given by
standard gap interpolation formula
$\Delta(t)=\Delta_0\tanh(2.2\sqrt{1/t-1})$,
where $\Delta_0$ is an energy gap at absolute zero.
The term of $F_1(-y)$ is given by the expression
$F_n(-y)=\int_0^{\infty}z^{n}(1+e^{z+y})^{-1}dz$.
Following the approach used for the specific heat~\cite{nakajima}, 
we generalize Eq.(\ref{eq.1}) to the two-gap model as
$\kappa_e=n_s\kappa_{e,s}+(1-n_s)\kappa_{e,l}$, 
where $\kappa_{e,i}$ ($i=l,s$) is the large and small gap bands thermal conductivity, respectively, and $n_s$ is a weight
for the small gap band.
In the following analysis, $n_s$ is fixed to be 0.6 regardless with the applied field, which is determined from the scaling.
To obtain $\kappa_n$, we estimate the phonon thermal conductivity
from the minima of $\kappa(H)/T$ shown in Fig.~\ref{fig.2}
assuming that the minima represent
at most the maximum value of the phonon contribution~\cite{mgb2thermal}.
% To achieve the semi-quantitative analysis,
% the estimation of the phonon thermal conductivity is required although it is not straightforward.
% Along with the way examined by Sologubenko {\it et al.}~\cite{mgb2thermal}, 
% we estimate the upper limit of the phonon thermal conductivity $\kappa_{\rm ph}/T$ above 1 K by assuming that the
% minima of $\kappa(H)/T$ shown in Fig.~\ref{fig.2} represent
% at most the maximum value of the phonon contribution.
% are caused by the competition
% of a decreasing $\kappa_{\rm ph}$ and an increasing $\kappa_{\rm e}$. 
% With this interpretation it is clear that the values of $\kappa_{\rm min}(H)/T$ 
% Black points shown in Fig.~\ref{fig.3} denote these minimum values of
% $\kappa(H)/T$ for different temperatures.
% On the other hand, at low temperature below 0.15 K, 
% we assume the temperature dependence of the phonon thermal conductivity $\kappa_{\rm ph}/T$ to be 0.11$\times T^{0.4}$
% as determined by the zero-field $\kappa/T$
% (the dashed line in Fig.~\ref{fig.3}).
% Following these assumptions, the total thermal conductivity in the superconducting state can be eventually expressed as 
% $\kappa=\kappa_e+\kappa_{\rm ph}=0.8\kappa_{e,s}+0.2\kappa_{e,l}+0.11\times T^{1.4}$.

\begin{figure}
\begin{center}
\includegraphics[scale =0.45]{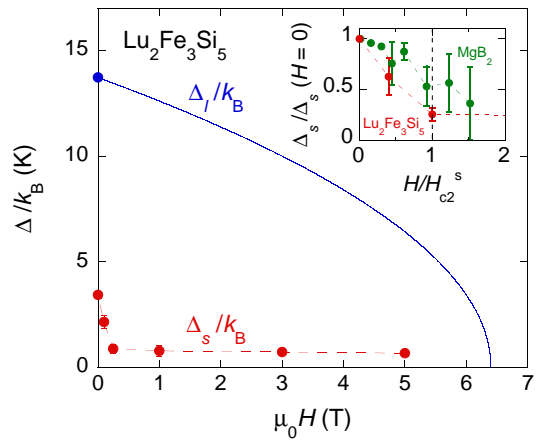}
\end{center}
\vspace{-0.5cm}
\caption{\label{fig.4} (color online). Field dependence of the superconducting 
gaps $\Delta/k_{\rm B}$. $\Delta_{l}/k_{\rm B}$ and $\Delta_{s}/k_{\rm B}$ represent the large and small gaps, respectively.
The solid line denotes the calculated result of $\Delta_{l}/k_{\rm B}$ based on the mean-field description.
The dashed line is a guide to the eye.
Inset: The normalized small gap $\Delta_{s}/\Delta_{s}(H=0)$ vs. $H/H_{c2}^s$ for Lu$_{2}$Fe$_{3}$Si$_{5}$ and
MgB$_2$~\cite{point}.}
\end{figure}

The solid lines shown in the inset of Fig.~\ref{fig.3} represent the calculated results of the total thermal conductivity $\kappa/T=\kappa_e/T+\kappa_{ph}/T$ 
based on the two-gap model at each field.
Here, we assume the temperature dependence of the phonon thermal conductivity to be $\kappa_{\rm ph}/T=0.11\times T^{0.8}$
as determined by the zero-field $\kappa/T$
(the dashed line in Fig.~\ref{fig.3}).
Moreover, at zero field, we use the gap values of $2\Delta_{0,l}/k_{\rm B}T_c=4.4$ and $2\Delta_{0,s}/k_{\rm B}T_c=1.1$
obtained from the specific heat measurement~\cite{nakajima}.
The field variation of $\Delta_{l}(H)$ is assumed to follow the 
mean-field description $\Delta_{l}(H)=\Delta_{0,l}\sqrt{1-H/H_{c2}}$ (the solid line in Fig.~\ref{fig.4}).
It should be emphasized that the two-gap model well reproduces the experimental results
especially for the maximum of $\kappa(T)/T$ only by tuning the small gap $\Delta_{s}/k_{\rm B}T_c$.
% We note that our analysis is not much affected by the estimation of $\kappa_{\rm ph}/T$.
The result further supports the existence of the small gap in Lu$_{2}$Fe$_{3}$Si$_{5}$.
Furthermore, the maximum most likely originates from an increase of the delocalized QPs excited over the small gap.
% In addition,  the light carriers in the small gap band further enhance the maximum.
% We should also claim that our analysis is not much affected by the estimation of $\kappa_{\rm ph}/T$ for $T>$ 1 K
% because the maximum associated with $\Delta_{s}$ appears below 1 K.
% For the $n_s$ parameter, 
% $n_s$ is determined to be 0.8 by the fitting to the data of $\kappa(T)/T$ at 5 T, and
% fixed to be constant regardless of the applied field.
On the other hand, 
there exists a discrepancy between the experiment and the calculation below the maximum temperature;
the experimental results show gradual decrease while the calculated $\kappa/T$ rapidly drops to zero.
One possible interpretation is that the system behaves like a dirty superconductor ($\xi > l_e$) 
due to the presence of $\mu_0H_{c2}^s$ that gives rise to a large coherence length $\xi$.
% of this difference is due to a small coherence length $\xi$ of the small gap.
It has been argued that the superconductors in the dirty regime shows
a rapid growth of QP density in the mixed state in comparison with those in the clean limit ($\xi \ll l_e$)~\cite{kusunose}.
% For the case of Lu$_{2}$Fe$_{3}$Si$_{5}$, the presence of the small $\mu_0H_{c2}^s$, and hence the large $\xi$ effectively makes the system 
% in the dirty regime.
% The effect of the finite concentration of impurities in our sample, which is not taken into account for the calculation,

The parameter of $\Delta_{s}/k_{\rm B}$ obtained from the analysis and $\Delta_{l}/k_{\rm B}$ are plotted against the magnetic field in Fig.~\ref{fig.4}.
It is clearly seen that $\Delta_{s}/k_{\rm B}$ sharply decreases at low field $\mu_0H_s \leq$ 0.25 T, while
$\Delta_{l}/k_{\rm B}$ shows monotonous decrease.
Note that the characteristic field scale of $\mu_0H_s$ is close to the virtual upper critical field $\mu_0H_{c2}^s \sim$ 0.26 T.
Interestingly, a similar field-induced suppression of $\Delta_{s}$ below $\mu_0H_{c2}^s$ is also observed in MgB$_2$
as demonstrated in the inset of Fig.~\ref{fig.4},
in which the gap values are determined by the point-contact study~\cite{point}.
For the case of MgB$_2$,
this behavior is understood as a consequence of a weak interband pairing interaction~\cite{koshelev,point}.
% is obtained from 
% and the numerical calculations for the $s$-wave multigap superconductivity 
% In fact, the strong field-induced suppression of $\Delta_{s}$ up to $\mu_0H_s$ is observed in both systems  
% In fact, the variation of the normalized small gap $\Delta(H)/\Delta(H = 0$ T) of Lu$_{2}$Fe$_{3}$Si$_{5}$ as a function of the normalized field $H/H_{c2}^s$
% shares a striking resemblance to
% that of MgB$_2$~\cite{point} as demonstrated in 
% as a model for the two-gap superconductor with a weak interband pairing interaction~\cite{koshelev}.
% Moreover, these observed behaviors are a common feature for the $s$-wave multigap superconductivity with a weak interband pairing interaction
% as discussed in MgB$_2$.
% In fact, the weakly coupling between the three-dimensional ($\pi$) band and the two-dimensional ($\sigma$)
% is thought to be an origin of the multigap superconductivity in MgB$_2$~\cite{Choi}. 
From the analogy of MgB$_2$,
the existence of  the multiple bands having the different dimensionality
and the weak interband interaction 
could be a source of the MGSC in Lu$_{2}$Fe$_{3}$Si$_{5}$.

In summary, our thermal conductivity measurements clarify the multigap superconductivity 
with a fully gapped excitation spectrum in Lu$_{2}$Fe$_{3}$Si$_{5}$.
The weak coupling between the distinct gap bands is thought to be the origin of the multigap superconductivity. 
In contrast to MgB$_2$, 
the dramatic enhancement of the quasiparticle heat conduction in the mixed state implies
the presence of the electronic correlations derived from the Fe 3$d$-electrons.
Our findings shed new light on the multigap superconductivity in the correlated electron systems.
% sharing with the iron-pnictides for
% understanding their intriguing superconducting properties.

We thank H. Harima, H. Kusunose, Y. Nakajima, and Y. Yanase for discussions.
This work is supported by Grant-in-Aids 
for Scientific Research  from JSPS and MEXT of Japan, and
a Grant-in-Aid for the Global COE Program of the Tokyo Institute of Technology.
% We thank H. Harima, H. Kusunose, Y. Nakajima, and Y. Yanase for discussions.
% This work is supported by Grant-in-Aids 
% for Scientific Research  from JSPS,
% a Grant-in-Aid for Scientific Research on
% Innovative Areas ``Heavy Electrons" from MEXT,
% and a Grant-in-Aid for the Global COE Program of the Tokyo Institute of Technology.

\bibliography{Lu2Fe3Si5v2}
\end{document}